\documentclass{emulateapj}
\usepackage{natbib}
\shorttitle{XMM-Newton Observation of Abell 3667 Relic}
\shortauthors{Finoguenov et al.}
\slugcomment{Draft version of \today}

\begin{document}

\title{XMM-Newton Observation of the Northwest Radio Relic Region in Abell 3667}

\author{%
Alexis Finoguenov\altaffilmark{1,2},
Craig L. Sarazin\altaffilmark{3},
Kazuhiro Nakazawa\altaffilmark{4},
Daniel R. Wik\altaffilmark{3},
and
Tracy E. Clarke\altaffilmark{5}}

\altaffiltext{1}{%
Max-Planck-Institut f\"ur extraterrestrische Physik, Giessenbachstra{\ss}e,
85748 Garching, Germany; alexis@xray.mpe.mpg.de}
\altaffiltext{2}{%
Center for Space Science Technology, University of Maryland Baltimore
County, 1000 Hilltop Circle, Baltimore, MD 21250, USA}
\altaffiltext{3}{%
Department of Astronomy, University of Virginia,
P. O. Box 400325, Charlottesville, VA 22904-4325, USA;
sarazin@virginia.edu; drw2x@Virginia.edu}
\altaffiltext{4}{%
Department of Physics, University of Tokyo, 7-3-1 Hongo,
Bunkyo-ku, Tokyo 113--0033, Japan; nakazawa@amaltha.phys.s.u-tokyo.jp}
\altaffiltext{5}{%
Naval Research Laboratory, 4555 Overlook Ave.\ SW,
Code 7213, Washington, DC 20375, USA; tracy.clarke@nrl.navy.mil}

\begin{abstract}
  Abell 3667 is the archetype of a merging cluster with radio relics. The NW
  radio relic is the brightest cluster relic or halo known, and is believed
  to be due to a strong merger shock.  We have observed the NW relic for
  $\sim 40$ ksec of net XMM time.  We observe a global decline of
  temperature across the relic from 6 to 1 keV, similar to the Suzaku
  results.  Our new observations reveal a sharp change of both temperature
  and surface brightness near the position of the relic.  The increased
  X-ray emission on the relic can be equivalently well described by either a
  thermal or nonthermal spectral model. The parameters of the thermal model
  are consistent with a Mach number ${\cal{M}}\sim2$ shock and a shock speed
  of $\sim$1200 km s$^{-1}$.  The energy content of the relativistic
  particles in the radio relic can be explained if they are (re)-accelerated
  by the shock with an efficiency of $\sim$0.2\%.. Comparing the limit on
  the inverse Compton X-ray emission with the measured radio synchrotron
  emission, we set a lower limit to the magnetic field in the relic of 3
  $\mu G$.  If the emission from the relic is non-thermal, this lower limit
  is in fact the required magnetic field.
\end{abstract}


\section{Introduction} \label{sec:intro}

Many clusters of galaxies appear to
be forming at the present time through massive cluster mergers.
Major cluster mergers are the most energetic events which have occurred
since the Big Bang, involving total energies of $\sim$$10^{64}$ ergs.
Merger shocks driven into the intracluster gas are the primary heating
mechanism of the gas in massive clusters.
Chandra and XMM-Newton
X-ray observations have provided beautiful images and spectra of merger
hydrodynamical effects, including cold fronts and merger bow shocks
\citep[e.g.,][]{Mar+00,MGD+02}.
However, the number of clusters with well-observed merger shocks is
limited.

\begin{figure*}
\includegraphics[width=17.cm]{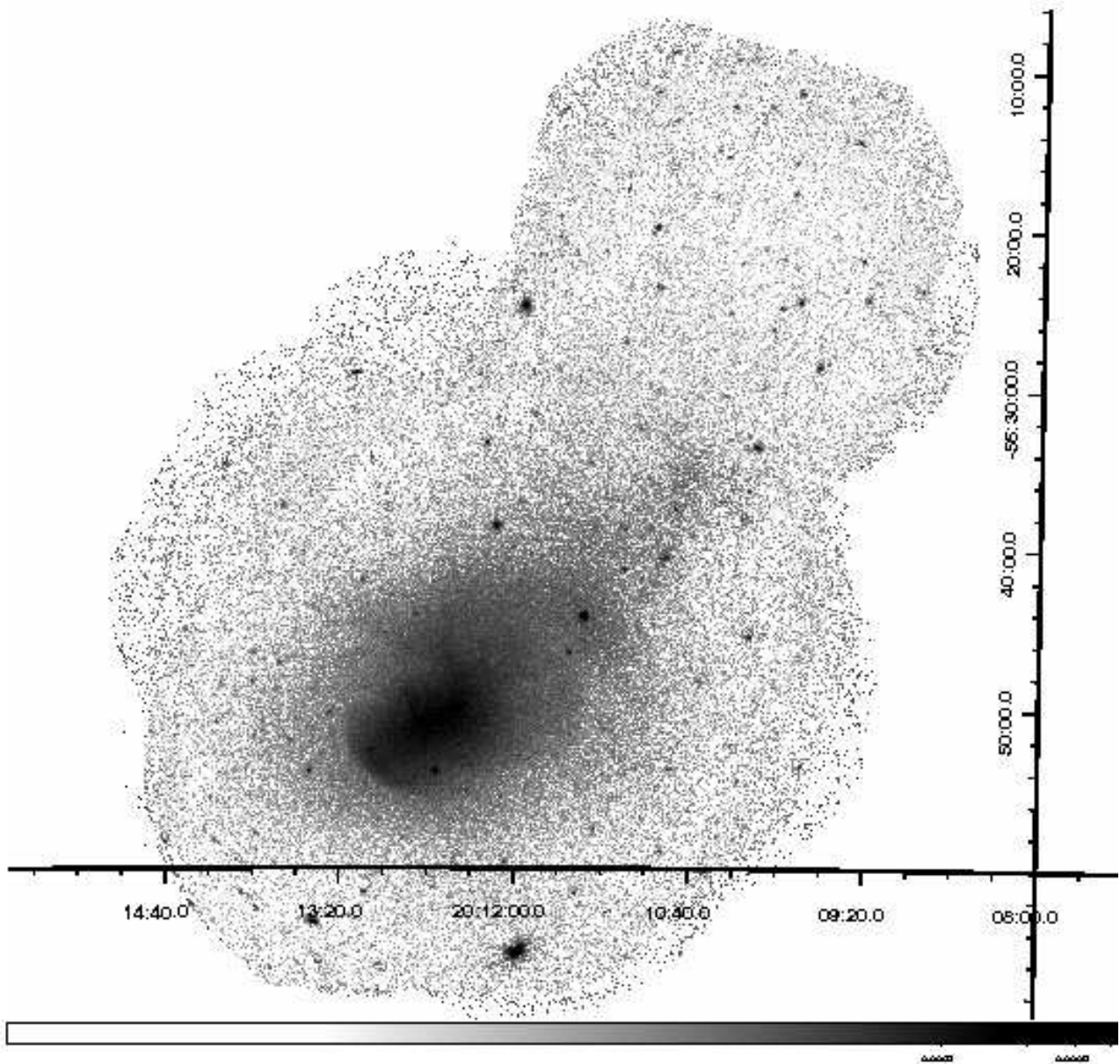}
\caption{XMM-Newton mosaic image of A3667 in the 0.5-2 keV energy band.
The gain in spatial coverage of the cluster due to the addition of our new northwest observation is evident.
The image has been background subtracted and exposure corrected.
\label{fig:mosaic}}
\end{figure*}

Enigmatic extended cluster radio sources with very steep spectra and no
clear
optical counterparts have been known for over 30 years
(\citealt{Wil70}, for a review see
\citealt{Fer08}).
These sources occur in a small fraction of rich galaxy clusters.  They have
such steep radio spectra that they can be detected only at lower frequencies
($\la$5 GHz).  Fairly symmetric sources which are projected on the cluster
center are often referred to as ``radio halos'' \citep[e.g., the Coma
cluster; ][]{DRL+97}, while similar elongated sources usually located on the
cluster periphery are called ``relics.''  The extremely steep radio spectra
flatten somewhat at low frequencies, which suggest that the relativistic
electrons have undergone significant losses due to synchrotron and inverse
Compton (IC) emission.

In every case, such diffuse cluster radio sources have been found in
irregular clusters which are apparently undergoing mergers.  This suggests
that the radio emitting electrons are accelerated or re-accelerated by
shocks or turbulence associated with these cluster mergers.  One possible
theoretical picture is that the radio halos are accelerated by turbulence
following the passage of merger shocks, while the relics are the direct
result of merger shock acceleration \citep[e.g.,][]{Fer08}.

The same relativistic electrons which produce the radio synchrotron
radiation will produce hard X-ray (HXR) emission by inverse Compton
scattering of Cosmic Microwave Background (CMB) photons.  However, the
detection of this emission has been difficult and remains controversial
\citep[e.g., compare][]{FOB+04,RM04}.

In many ways, Abell 3667 is the ideal site to study mergers and radio
relics.  It is a very bright X-ray cluster at a low redshift ($z = 0.0552$).
The ROSAT and ASCA observations showed that it is a spectacular merger with
shock heated gas \citep{MSV99}.  The galaxy spatial and redshift
distributions are also strongly bimodal \citep{OCN09}.  A3667 was one of the
first clusters in which a merger cold front was found with Chandra
\citep{VMM01a,VMM01b}, and the term was coined based on the A3667
observations.  The very sharp density and temperature gradients in this cold
front have provided very strong limits on the role of transport processes
(like heat conduction) in clusters.  Subsequent Chandra and XMM/Newton
observations \citep{MFV02,BFH04} have provided spectacularly detailed images
and information on the dynamics of the merger.  However, most of the Chandra
and XMM observations study the interior regions of the cluster, not the
outer regions near the radio relics.

Abell 3667 contains a pair of curved cluster radio relics \citep{RWH+97}.
Their location on either side of the cluster center at 1.8Mpc distance
($29^\prime$) and sharp, inwardly curved outer edges are exactly what is
expected for merger shocks and shock particle acceleration.  Models which
reproduce the optical and inner X-ray properties of A3667 predict shocks at
or near these locations \citep{RBS99,RS01}.  The radio spectra steepen with
distance from the outer edge \citep{RWH+97} as expected if the electrons are
accelerated there, and the higher energy electrons lose energy due to
synchrotron and IC emission as they are advected away from the shock.

The northwest radio relic in Abell 3667 is the brightest (highest flux)
cluster radio relic or halo source which is known, with a flux at 20 cm of
3.7 Jy \citep{Joh-Hol04}.  It is also one of the largest relics, with a
total extent of 33\arcmin\ or 2.1 Mpc. Since the electrons which produce the
IC HXR emission are basically the same ones which emit radio synchrotron,
for a given magnetic field the HXR flux should be nearly proportional to the
radio flux.  Thus, Abell 3667 might be expected to be the brightest
nonthermal (NT) cluster hard X-ray source.  Since this relic is at a large
projected radius from the cluster center, if anything the magnetic field
should be lower than in other objects, implying an even larger HXR flux.

\begin{figure*}
\includegraphics[width=17.cm]{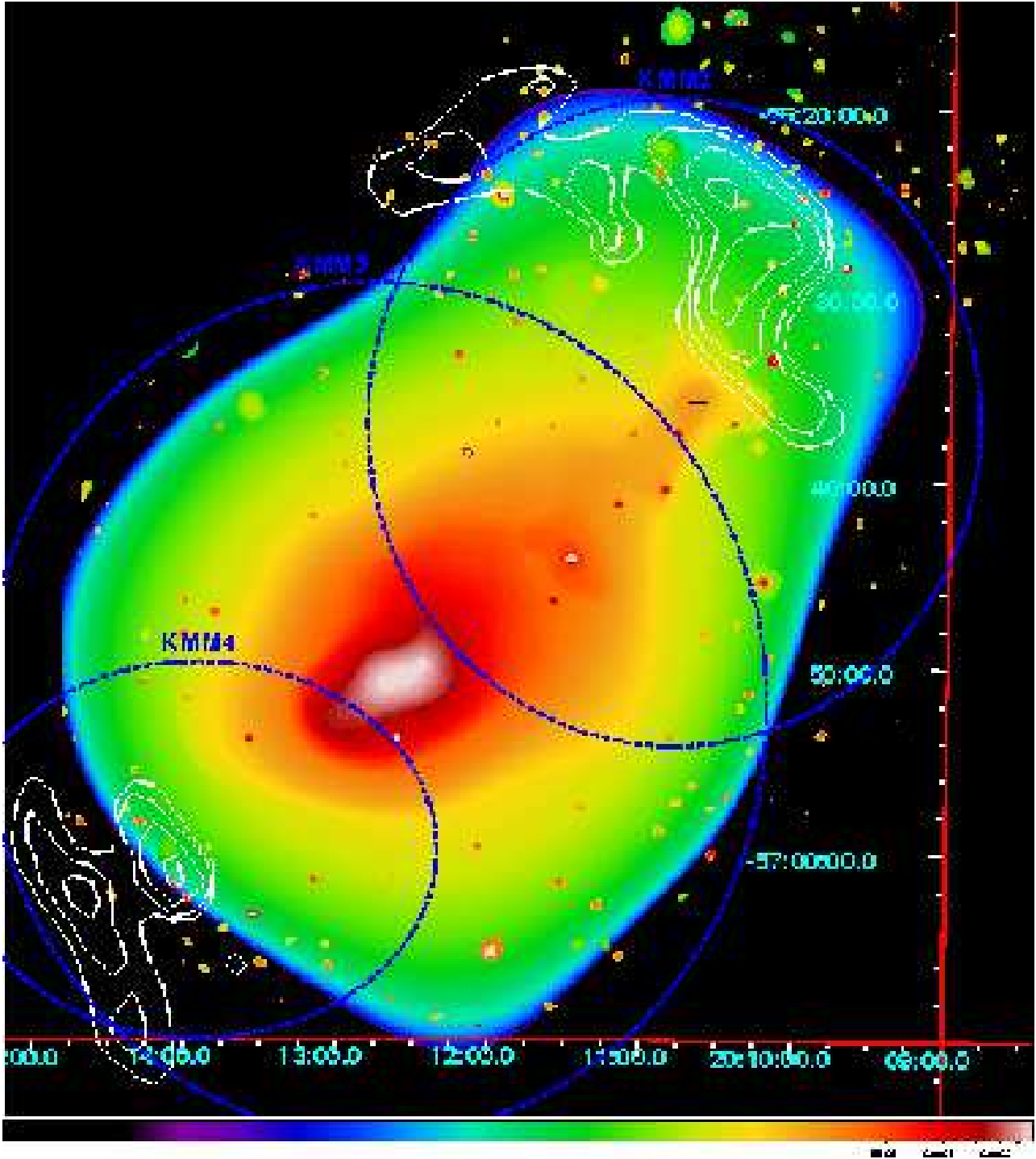}
\caption{Wavelet reconstruction of the 0.5-2 keV X-ray mosaic of A3667 from XMM.
There is a sharp discontinuity in the X-ray surface brightness to the
northwest;  The blue ellipses show the location, name and size of three main
galaxy components (KMM5, KMM2 and KMM4) of A3667 from the analysis of Owers
et al. (2009). The black cross marks the center of KMM2 group.
The white contours show the wavelet reconstruction of the SUMMS 843 MHz
radio image in the region of radio relics.  The contours are drawn at the
1, 3, 9 and 20 mJy/beam levels. 
\label{fig:mosaic-wv}}
\end{figure*}

We have a long Suzaku observation of the NW radio relic region of Abell 3667
with Suzaku \citep{Nak+09}, as well as two observations of more central
regions of the cluster.  While hard X-ray emission is seen coming from the
cluster center with HXD intrument \citep{Nak+09}, it could be thermal
emission, while there is no clear Suzaku detection of inverse Compton hard
X-ray emission from the NW radio relic, given the systematic uncertainties
in the non-X-ray background and thermal emission from the cluster.  Our
Suzaku observation also provides a CCD X-ray image of a portion of the radio
relic region with the X-ray Imaging Spectrometer (XIS).  However, this
covers only a part of the region of the radio relic and possible merger
shock.  

Here, we present the results of a new XMM-Newton observation of Abell~3667
which is centered on the NW radio relic.  We assume a concordance cosmology
with $H_0 = 71$ km s$^{-1}$ Mpc$^{-1}$, $\Omega_M=0.27$, and
$\Omega_\lambda=0.73$.  At the redshift of Abell 3667 of $z=0.05525$, the
angular diameter distance is $d_A = 212.6$ Mpc, and 1\arcmin\ corresponds to
61.8 kpc.  For a mean cluster temperature of 7.2 keV, the virial radius
should be $r_{180} \approx 2.33$ Mpc $= 38\arcmin$ \citep{MFS+98}. Unless
otherwise specified, we provide confidence intervals at the 68\% level.

\section{Observation and Data Analysis} \label{sec:data}

There have been seven previous XMM-Newton observations of Abell 3667.  Six
of these were discussed in \citet{BFH04}, where these observations were
merged to make a mosaic of the center of the cluster.  There was a
subsequent observation (OBSID 0206850101) of the center of the cluster,
which we analyzed and added to the mosaic for completeness.  All these
observations, however, cover only the central part of the cluster.  In order
to study the characteristics of the X-ray emission at the location of radio
relic, we proposed for an additional XMM-Newton observation, with the
pointing direction selected to ensure a robust background subtraction can be
performed by leaving enough area beyond the location of relic.  The
observational ID of this pointing is 0553180101, and it was performed during
XMM-Newton orbit 1620.  In this paper, we will concentrate on the results
obtained only from this observation.  The total scheduled time of
observation was 55 ksec.  We used XMMSAS (Watson et al.\ 2001) version 7.3
for the standard data reduction.  After producing the event files, we have
applied a strict light curve cleaning in order to ensure that we retained
only the periods with the lowest and the most stable background.  The
resulting net exposures after the cleaning are 37, 47 and 44 ksec for the
pn, MOS1, and MOS2 detectors, respectively.  The medium filter has been used
for all instruments, for consistency with the previous observations of the
cluster. Extended full frame mode has been used for the pn to reduce the
effect of out-of-time events.  In producing the mosaic, we have checked
whether any of the ccds exhibit unusually high background, identifying none.

\section{X-ray Image Analysis} \label{sec:image}

We used the quadruple background subtraction technique of \citet{Fin+07} to
produce the mosaic image of A3667.  Due to high level of cluster emission,
several iterations have been performed in order to locate and mask out the
bright parts of the cluster.  We used both pn and MOS data in producing the
mosaic image.  Figure~\ref{fig:mosaic} shows the mosaic image made with the
new observation and the previous data from XMM in the 0.5--2 keV band.  It
has been corrected for background, exposure, vignetting, and out-of-time
events, and excluded the 1.4--1.6 keV energy band due to the strong
background Al line.  Comparison to previous XMM X-ray images
\citep{BFH04,Nak+09} shows the significant gain in spatial coverage achieved
with the new observation.  The image in Figure~\ref{fig:mosaic-wv} shows a
corresponding wavelet reconstruction using the method of \citet{VMF+98} and
applying the image reconstruction technique of Finoguenov et al.\ (2009) to
remove the wings of point sources.

To study the spatial correlation of the X-ray emission to the radio relic,
we used the SUMMS radio data.  SUMSS is a deep radio survey of the entire
sky south of declination $-30$ degrees, made using the Molonglo Observatory
Synthesis Telescope, operating at 843 MHz and recording right-circular
polarization \citep{BLS99}.  The beam size of these radio images is
45\arcsec.  We used the data which are a part of SUMSS release 0.2.  To
characterize the radio relics, we apply the wavelet reconstruction of radio
image, remove point-like features and display the contours at the 1, 3, 9
and 20 mJy/beam levels in Figure~\ref{fig:mosaic-wv}.  The wavelet
reconstruction of the XMM image of Abell 3667 shows an interesting
coincidence between the extent of the significant X-ray emission in the
0.5--2 keV detected on the wavelet scales upto $8^{\prime}$ and the outer
edge of the radio relic.  The detection stops $11^{\prime}$ from the end of
the data, compared to $2^{\prime}$ on the opposite edge of the cluster. As
discussed in \S~\ref{sec:intro}, this is the expected location of the
cluster merger shock, if the relic was produced by shock acceleration or
re-acceleration of relativistic electrons.

%

Our previous Suzaku XIS image of this region did not show this edge
\citep{Nak+09}, which we believe to be a result of contamination by point
sources which were not individually resolved due to the large Suzaku XIS
point spread function (PSF).  Many of these point sources are seen in
Figures~\ref{fig:mosaic} \& \ref{fig:mosaic-wv}.  In addition there is a
steep gradient in the cluster emission to the southeast, associated with the
major subcluster, which  is smeared by the large Suzaku XIS PSF and
covers the relic.  
To test the effects of these features and the Suzaku XIS PSF on the Suzaku
image, and the consistency of the Suzaku and XMM data, we blurred the XMM
image using a Gaussian with a sigma of 1\arcmin, which is a crude
approximation to the Suzaku XIS PSF. We found that this blurred image
matched the actual Suzaku XIS image fairly well, and the surface brightness
discontinuity observed with XMM at its full resolution is no longer apparent
in the smeared image.

A more detailed examination of the XMM image shows in addition to the X-ray
peak associated with the main cluster, another peak, with coordinates of the
X-ray center are RA $= 20^h 10^m 41\fs285$, Decl. $ = -56\arcdeg 35\arcmin
45\farcs81$ (J2000).  The available results of the spectroscopic survey of
the cluster \citep{JHC08, OCN09} allows us to identify this peak with the
major subcluster in A2667, KMM2, found by \citep{OCN09}, which plays a key
role in the dynamics of the system.  In fact the relics are not symmetric
with respect to any of the galaxy component but the region between KMM2 and
KMM5, indicating that within the forward shock hypothesis of their origin,
it is the recent passage of KMM2 through KMM5 that is responsible for their
appearence.

\includegraphics[width=8.cm]{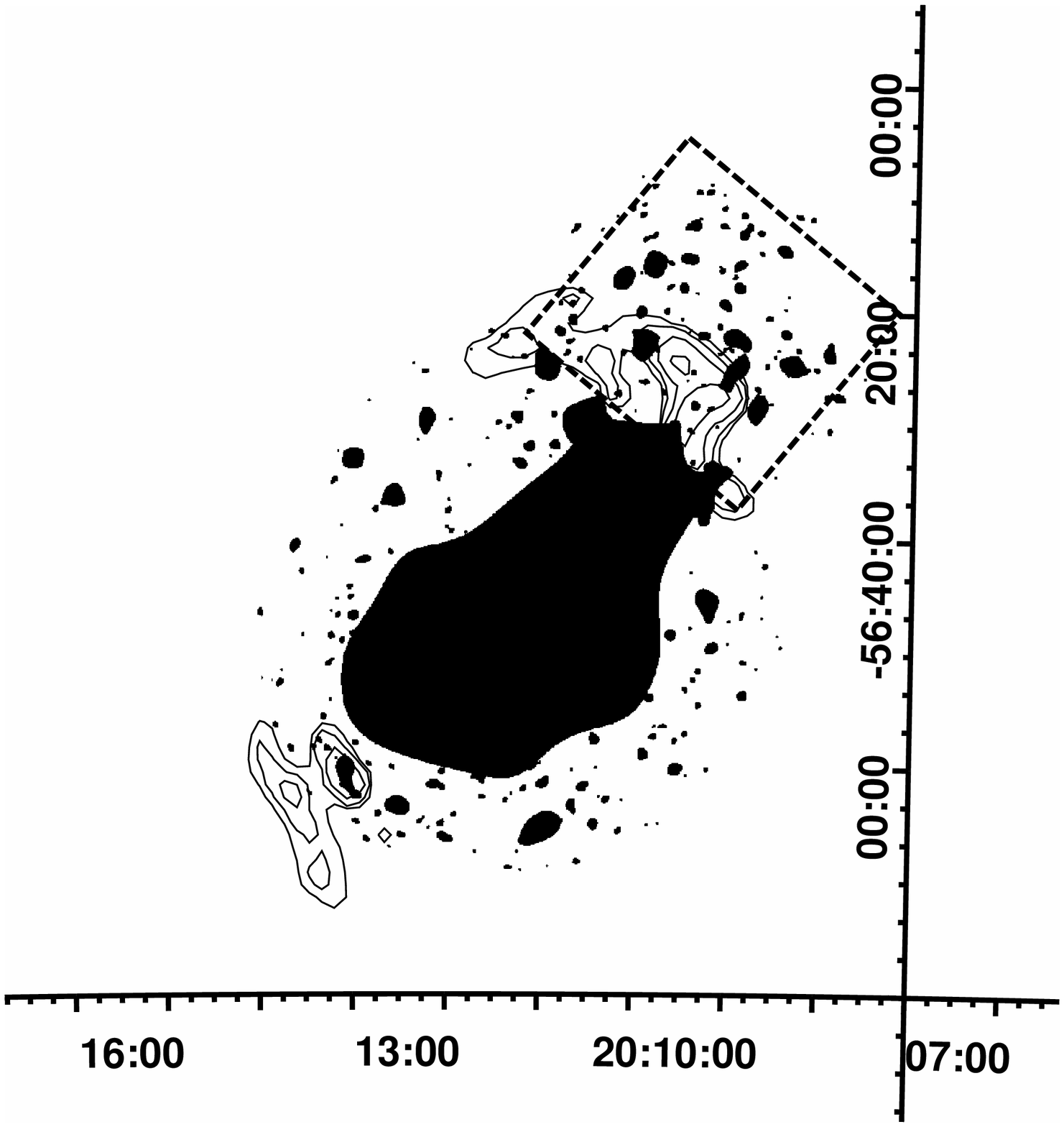}
\figcaption{The zones masked out from both the surface brightness and
  spectral analysis.  We removed all significant point sources, as well as
  the bulk of the cluster emission and the emission from the two northwest
  galaxy groups. The dashed box shows the region for surface brightness
  extraction presented in Fig.\ref{fig:sb}. The black contours show the
  wavelet reconstruction of the SUMMS 843 MHz radio image in the region of
  radio relics.  The contours are drawn at the 1, 3, 9 and 20 mJy/beam
  levels.
\label{fig:mask}}

In order to avoid biased results caused by the possibly different
temperature gas associated with this subcluster, we excluded the region from
further analysis of both the cluster and relic emission, together with all
the point and compact extended sources detected in the image. We used the
wavelet images on spatial scales of 4, 8, 16, 32, and 64 arcsec to mask out
such regions.  Figure~\ref{fig:mask} shows all of the regions masked out
from the surface brightness and spectral analysis.

In Figure~\ref{fig:sb}, we show the X-ray surface brightness radial profile
around the radio relic.  We used the 1.6-4.0 keV band to extract the profile
and use only pn data. The low-energy cut off has been selected to avoid any
possible influence of the strong low-energy emission seen towards A3667
\citep{BFH04}, which is likely of Galactic origin. The upper energy is
selected to maximize the signal-to-noise, as at higher energies the
background dominates.  The background in the selected energy band is
dominated by particle background which is not vignetted, and thus is roughly
constant in surface brightness.
We see a sharp drop in the surface brightness profile at the position of the
outer edge of the relic, indicating that there is an association of this
X-ray edge with the radio relic. A small peak in the radio surface
brightness at $+10^\prime$ distance from the relic is accosiated with two
radio point sources. We associate a corresponding low-significance
enhancement in X-rays is two sources below the detection threshold.

\includegraphics[width=8.cm]{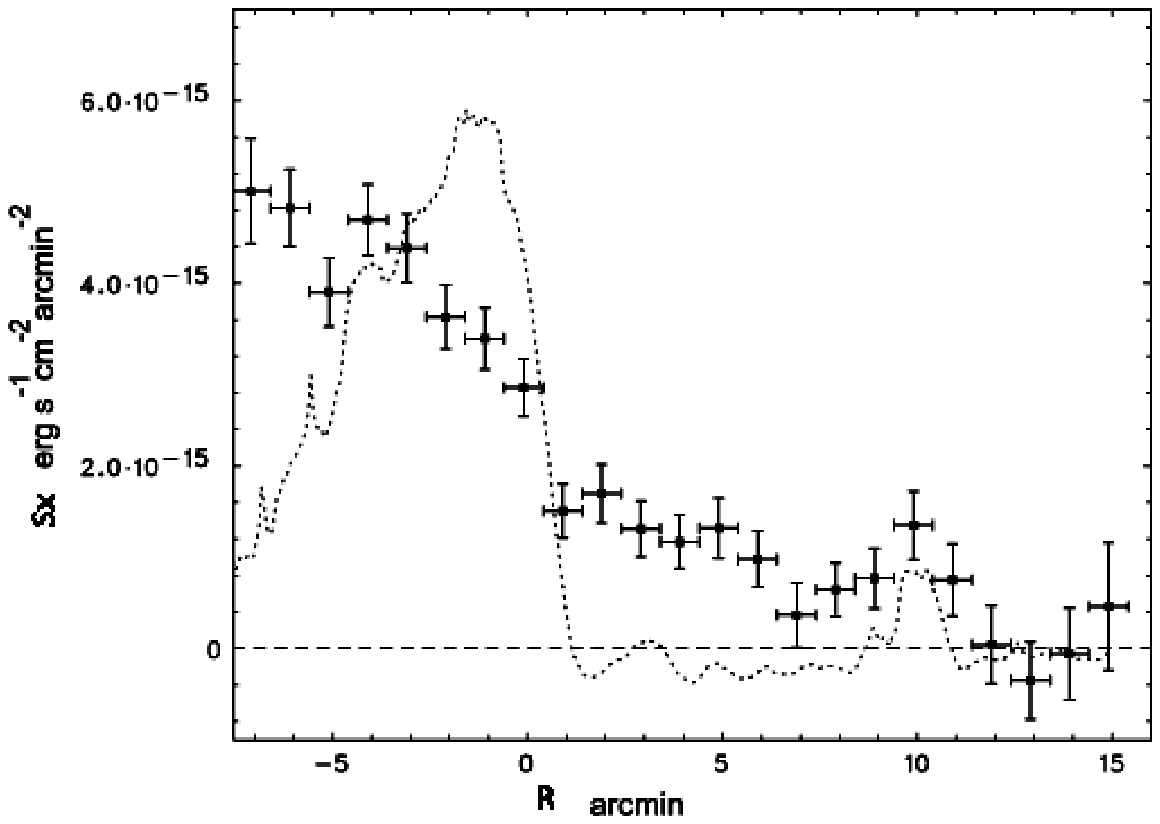}
\figcaption{The surface brightness profile of A3667 in the 1.6-4.0 keV band
  extracted along the direction from the cluster center (negative distance
  to the relic) to the northwest.  The radial units are arcminutes.  The
  location of the outer edge of the relic is at 0. Point sources were
  excised.  The dashed line shows the surface brightness profile of radio
  emission, extracted in the same fashion from the 843 MHz SUMMS data.  One
  sees a sharp drop in the surface brightness at the position of the relic
  and a much flatter behaviour both at small and at large X-axis values.
\label{fig:sb}}

\begin{figure*}              
\includegraphics[width=16.cm]{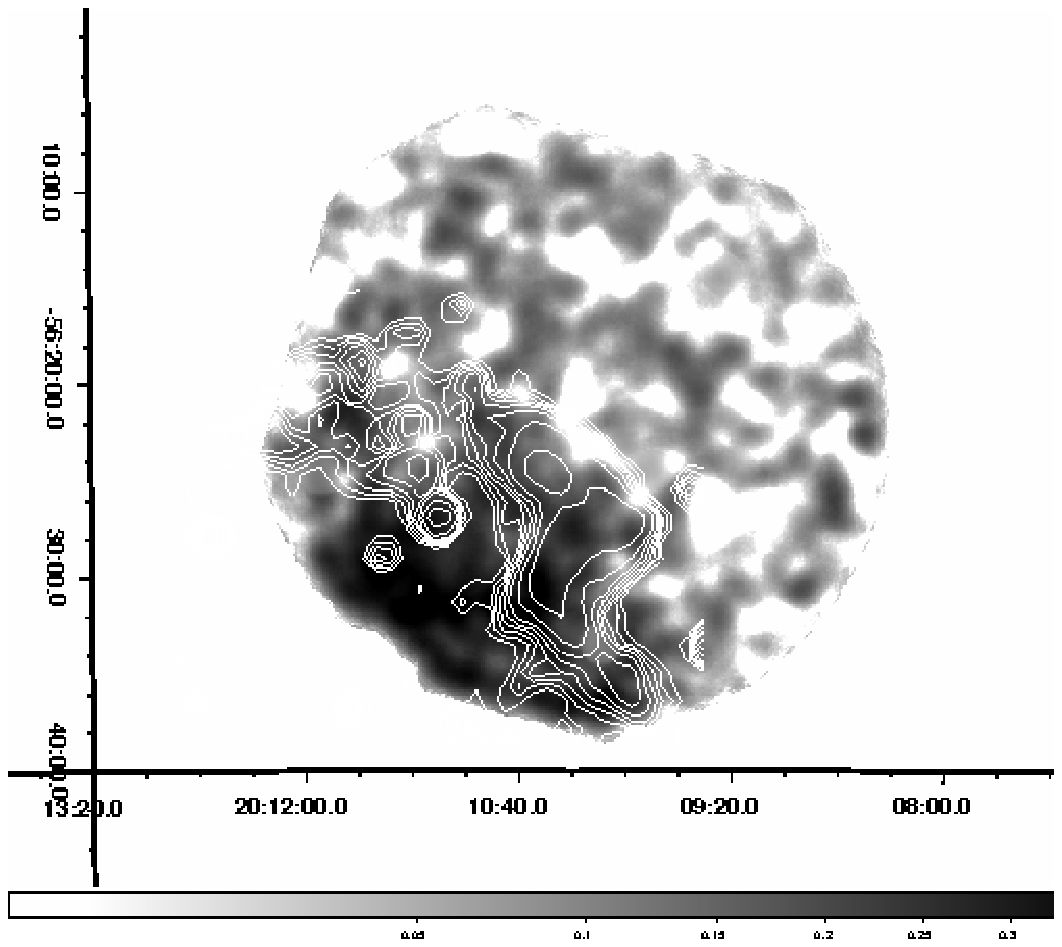}
\figcaption{XMM-Newton image (pn+MOS) of the region around the NW radio
  relic in A3667 in the 1.6-4 keV energy band.  The X-ray point sources have
  been removed.  The image was smoothed with a 32\arcsec\ Gaussian.  The
  white contours show the radio emission without the wavelet reconstruction
  applied.  Dark zones indicate the regions of intense X-ray emission.  The
  image shows that the complex shape of the radio emission is followed by
  the X-rays.
\label{fig:relic-wv}}
\end{figure*}

In Figure \ref{fig:relic-wv}, we show the combined pn+MOS XMM-Newton image
in the 1.6-4 keV band for the region immediately around the radio relic.
There is a clear association of the surface brightness discontinuity with
the location of the outer edge of radio relic. This image was not corrected
for vignetting to show that the result is not affected by the details of
background subtraction (vignetting correction in combination with
undersubtraction of the background can lead to enhancements at the edge of
the detector). The image is smoothed with a 32\arcsec\ Gaussian after the
removal of point sources. The X-ray image appears to follow the radio relic
in some detail.  In particular, the X-ray surface brightness discontinuity
traces the outer, sharp edge of the radio relic, including the indentation
at the west end of the relic. To quantify the trends seen in Figure
\ref{fig:relic-wv}, we proceed with the spectral analysis.

\begin{deluxetable*}{lccccccl}
\tabletypesize{\footnotesize}
\tablewidth{0pt}
\tablecolumns{8}
\tablecaption{Thermal Model Spectral Fits to the X-ray Emission near the Relic\label{tab:spect_t}}
\tablehead{
\colhead{} &
\colhead{$kT$} &
\colhead{Norm.\tablenotemark{a}} &
\colhead{} &
\colhead{} &
\colhead{Area} &
\colhead{$I_X$ (2--5 keV)} &
\colhead{} \\
\colhead{Zone ID} &
\colhead{(keV)} &
\colhead{($10^{-5}$)} &
\colhead{$\chi^2_r$} &
\colhead{$N_{d.o.f.}$} &
\colhead{(arcmin$^2$)} &
\colhead{($10^{-15}$ ergs/s/cm$^2$/arcmin$^2$)} &
\colhead{Comment}
}
\startdata
1 & $1.0\pm0.2$ & $1.0\pm0.2$ & 0.90 & 297       & 60.4 & $0.9\pm0.6$   &background - main \\
2 & $4.0\pm0.8$ & $17.5\pm1.7$ & 0.90 & 111      & 23.3 & $8.9\pm1.0$   & relic back-1\\
3 & $4.7^{+2.2}_{-0.8}$& $15\pm1.5$ &0.82 & 75 & 21.8 & $13.2\pm1.8$  & relic back-2 \\
4 & $5.1^{+4.1}_{-1.6}$& $9.1\pm 1.1$ &1.06&149& 102.1& $4.3\pm0.5$   & relic front\\
5 & $5.4^{+3.3}_{-1.5}$& $6.0\pm0.6$ &1.33 & 54& 28.0 & $8.0\pm1.3$   & relic back-3 \\
6 & $3.8\pm0.8$ & $24\pm 1.4$ & 0.96 & 131     & 80.8 & $11.4\pm1.2$  & cluster near relic \\
7 & $6.0\pm1.3$ & $26\pm 1.7$ & 0.89 & 91      & 63.2 & $19.2\pm1.8$  & cluster near relic \\
8 & $4.7\pm1.2$ & $24\pm 1.7$ & 1.10 & 100     & 82.4 & $20.4\pm2.1$  & cluster near relic \\
9 & $1.0\pm0.1$ & $4.1\pm 0.5$ & 1.19 & 49     & 56.9 & $2.2\pm0.8$   & zone in front of relic \\
10 & $1.9\pm0.6$ & $12\pm 2$ & 1.03 & 441      & 765.2& $0.9\pm0.3$   & zone in front of relic \\
11 & $1.7\pm0.8$ & $6.9\pm 1.7$ & 1.06 & 137   & 259.2& $0.6\pm1.0$   & zone in front of relic \\
12 & $2.5\pm0.5$ & $37\pm 2.2$ & 1.05 & 290    & 503.6& $4.3\pm0.7$   & eastern relic \\
13 & $ 1.8\pm 0.6$ & $15\pm 2$ & 0.91 & 245    & 476.6& $1.7\pm0.5$   & zone west of relic \\
14 & $1.0\pm0.3$ & $0.12\pm0.05$  & 0.92 & 136   & 340.3& $2.2\pm1.2$   & background side \\
15 & $ 1.6 \pm 0.4$ & $4.8\pm 2$ & 1.38 & 49   & 75.7 & $6.0\pm1.5$   & zone behind the relic \\
\enddata
\tablenotetext{a}{Normalization of the APEC thermal spectrum,
which is given by $\{ 10^{-14} / [ 4 \pi (1+z)^2 d_A^2 ] \} \, \int n_e n_H
\, dV$, where $z$ is the redshift, $d_A$ is the angular diameter distance,
$n_e$ is the electron density, $n_H$ is the ionized hydrogen density,
and $V$ is the volume of the region.}
\end{deluxetable*}


\begin{deluxetable}{cccccc}
\tabletypesize{\footnotesize}
\tablewidth{0pt}
\tablecolumns{6}
\tablecaption{Power-Law Spectral Fits to the Relic X-ray Emission\label{tab:spect_pl}}
\tablehead{
\colhead{Zone ID} &
\colhead{$\Gamma$} &
\colhead{Norm.\tablenotemark{a}} &
\colhead{$\chi^2_r$} &
\colhead{$N_{d.o.f.}$} &
\colhead{comment}
}
\startdata
2 &$1.6\pm0.1$ & $15\pm4$ & 0.91 & 111 & relic back-1\\
3 &$1.8\pm0.1$ & $38\pm2$ & 0.79 & 75   & relic back-2 \\
4 &$1.4\pm0.2$ & $17\pm3$ & 1.09 & 149   & relic front\\
5 &$1.9\pm0.2$ & $16\pm2$ & 1.28 & 54   & relic back-3 \\
\enddata
\tablenotetext{a}{The normalization of the power-law is the
photon flux at 1 keV in units of
$10^{-6}$ photons cm$^{-2}$ s$^{-1}$ keV$^{-1}$.}
\end{deluxetable}

\section{Spectral Analysis} \label{sec:spectra}

In order to investigate the origin of the break in the surface brightness,
we extracted the X-ray spectra from several regions around the radio relic.
The regions were designed to encompass the zone directly in front of the
relic, much of the relic itself, as well as other regions tracing the
possible azimuthal variations, and cluster radial temperature profile.
These spectral extraction regions are shown in Figure~\ref{fig:regions}.
We used the spectral range of 0.4-14 keV in the spectral analysis.  We
started by fitting the most distant zone (from the cluster center) covered
by XMM observation.  The energies above 7 keV where clearly dominated by the
residual particle background, which we fit with a background (not convolved
with the effective area of the telescope) power law model. The residual
emission was found to consists of a $kT \sim 0.3$ keV thermal emission we
associate with the Galactic foreground emisson and a significant 1 keV
component, consistent with the Suzaku XIS results, which we therefore
associate with the A3667 cluster emission. The outmost zone is also the
largest zone and the constraints on the background components are the best
there, so we fixed the slope of the particle background and the temperature
of the soft emission to the values found there, while the normalization of
both these components was left free to vary. We found this
background/foreground model to be a satisfactory fit, and the reported
additional thermal and non-thermal components we associate with the cluster
are clearly in excess of it.  Since no spatial variation in the particle
background spectral shape has been reported [except for the case of bright
MOS chips (Kunz \& Snowden 2008), which we associated with electronic noise
and not with particle background itself], we are certain
that the reported results correspond to a detection of a change in the
intrinsic properties of the cluster emission.

\includegraphics[width=8.cm]{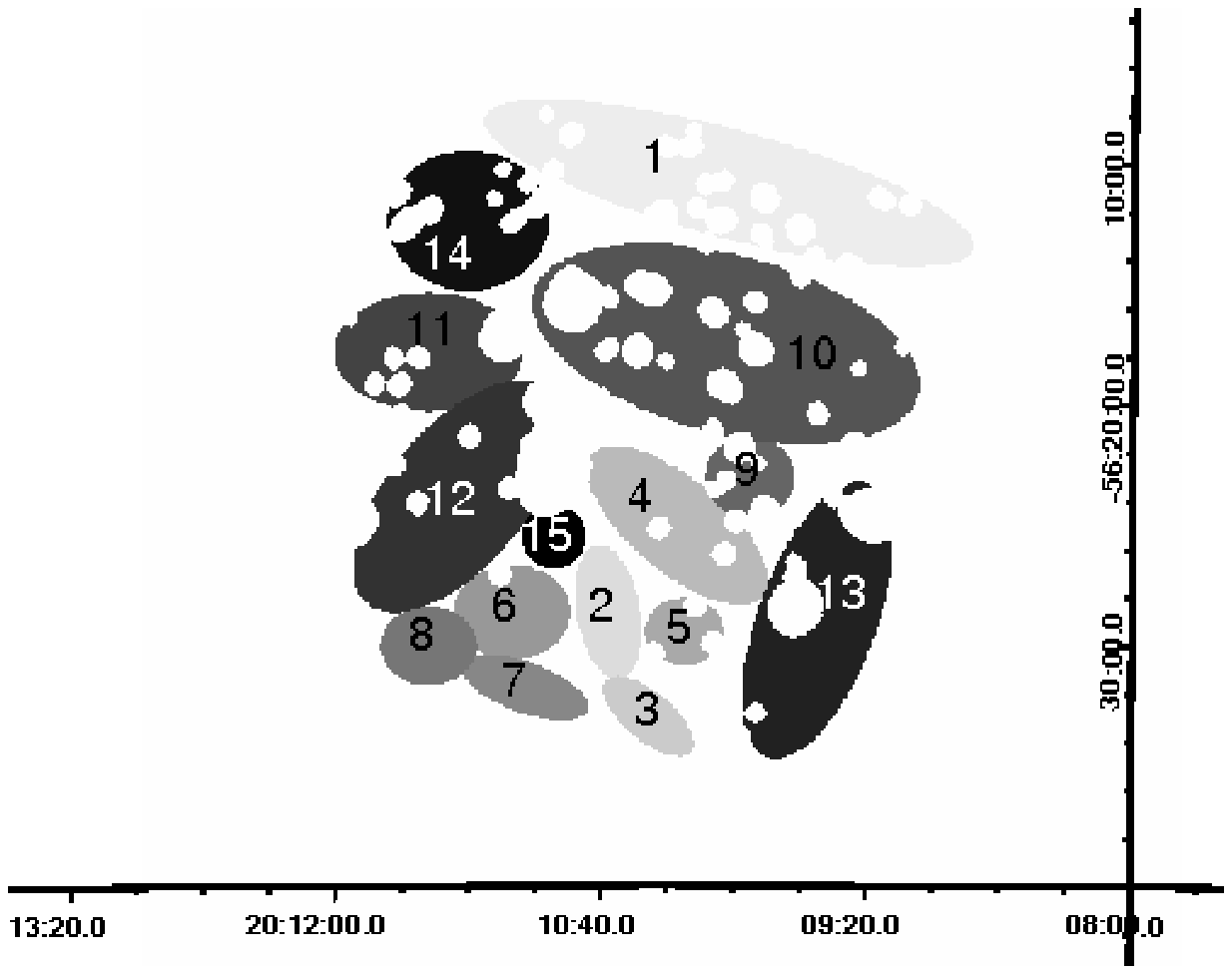}
\figcaption{The spectral extraction regions for the NW  radio relic.
The numbers correspond to the Zone IDs in Tables~\ref{tab:spect_t} \& \ref{tab:spect_pl}.
\label{fig:regions}}

The results for the spectral fits for the cluster emission components are
given in Tables~\ref{tab:spect_t} \& \ref{tab:spect_pl}.  The first table
gives the result of purely thermal spectral fits using APEC plasma code
(Smith et al.\ 2001), while the fits in Table~\ref{tab:spect_pl} include
nonthermal power-law components.

We first consider the purely thermal fits to the spectra.  The results are
summarized in Table~\ref{tab:spect_t}, where the columns give the Zone ID
from Figure~\ref{fig:regions}, the temperature $T$, the normalization of the
APEC thermal model, the reduced chi-square $\chi^2_r$, the number of degrees
of freedom in the fit $N_{d.o.f.}$, the projected solid angle of the region
excluding the area masked to avoid point sources, the average X-ray surface
brightness of the region $S_X$ in the 1.6-4.0 keV band, and a comment on the
reason for choosing this region.
The element abundance has been set to 0.2 solar (typical for cluster
outskirts), as no strong lines have been seen in the emission to constrain
it.

\includegraphics[width=8.cm]{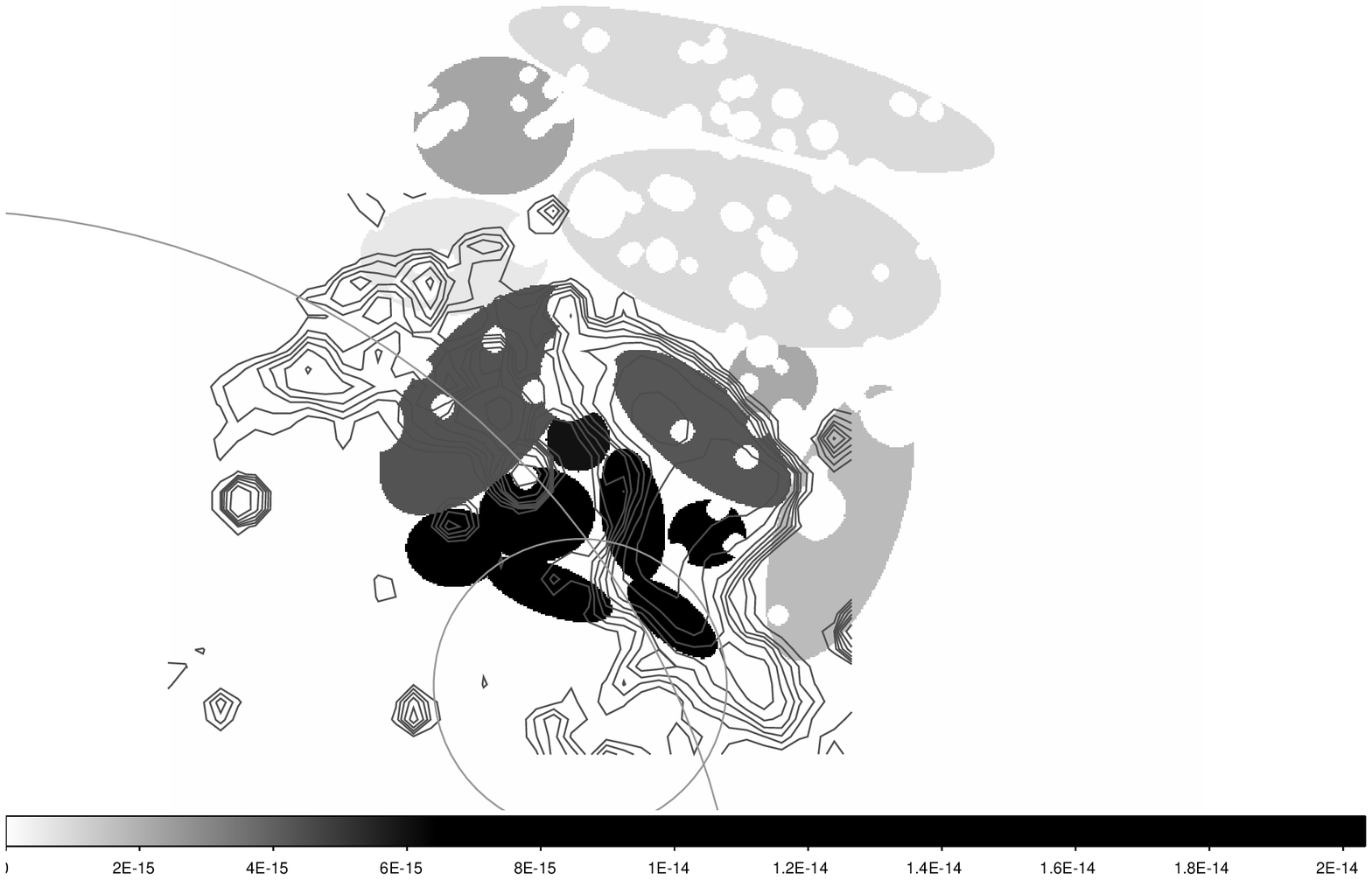}
\figcaption{The reconstructed surface brightness in the 2--5 keV band using
  the results of the spectral fits, and directly fitting for the spectral
  shape of the background without assuming a flat spatial distribution.  The
  grey scale shows the flux level in units of ergs s$^{-1}$ cm$^{-2}$
  arcmin$^{-2}$.  We note that an enhancement in surface brightness is
  located on the relic, as seen by the difference in regions on either side
  of the relic. The larger circle shown is centered on the cluster center.
  The smaller circle is centered on the main subcluster.
  \label{fig:sb2050}}

As one moves inward in radius towards the relic, there is initially a slow
rise in the temperature, from 1 to 2 keV, followed by an abrupt increase in
the temperature from $kT \sim 1.9$ keV in front of the relic to $ \sim 5$
keV on the relic.  Figure~\ref{fig:sb2050} shows the average X-ray surface
brightnesses of each of the regions plotted along with contours showing the
radio relic. We confirm the dent-like structure of the X-ray emission, seen
in Fig.\ref{fig:relic-wv}.  Figures~\ref{fig:sb} \& \ref{fig:sb2050} both
show that there is also a jump in the X-ray surface brightness inward of the
outer edge of the radio relic. The value of the jump is higher in
Figure~\ref{fig:sb2050}, compared to Figure~\ref{fig:sb}, which we attribute
to the presence of the dent in the X-ray emission, which was separated
though the selection of spectral extraction zones, but not in the extraction
of the surface brightness profile.

The combination of the jump in the temperature and the
jump in the surface brightness (which is determined mainly by the gas
density) are consistent with a location of the shock at or near the front of
the relic. The parameters of the shock are discussed in the next section.
In any case, whatever its origin, the hardening of the spectrum at the front
of the relic is significant at better than the 95\% confidence level.


Alternatively, the jump in the X-ray surface brightness and spectral
hardening might also be attributed to the non-thermal emission associated
with the relic.  To address this possibility, we also fit the spectra in the
regions corresponding to the projected position of the NW radio relic with a
non-thermal power-law spectrum.  The results of these fits are summarized in
Table~\ref{tab:spect_pl}, where the columns give the Zone ID from
Figure~\ref{fig:regions}, the power-law photon spectral index $\Gamma$, the
normalization of the power-law which is the flux at 1 keV, the reduced
chi-square $\chi^2_r$, the number of degrees of freedom in the fit
$N_{d.o.f.}$, and the same comment on the location of the region as given in
Table~\ref{tab:spect_t}.  Note that the photon spectral index values are
generally lower than might have been expected from the steep radio spectrum
of the radio relic.  The average radio spectral index between 85 MHz and 2.3
GHz for the whole relic is $\alpha \approx 1.1$ \citep{RWH+97}, which
corresponds to $\Gamma \approx 2.1$.  However, the fits in
Table~\ref{tab:spect_pl} are for pure power-law emission.  Almost certainly,
there is also some thermal emission from the intracluster gas, which may
affect the fitted spectral index of the power-law.
The limited statistics in the observed X-ray spectra of these regions do not
allow for more complicated thermal plus power-law models.  Also, for
reasonable values of the magnetic field, the X-ray emission in the
relatively soft XMM band would come from lower energy relativistic electrons
than produce most of the observed radio emission. The curvature in the
observed radio spectrum indicates that the electron energy spectrum flattens
at lower energies.  A steep radio spectrum at high frequencies is expected
as a consequence of the radiative losses by the relativistic electrons.

Unfortunately, as a result of the limited statistics in the spectra of these
faint regions in the outer parts of this cluster, it isn't possible at this
time to determine whether thermal or non-thermal models provide a better fit
to the data. If one considers zone 4 as an example, the thermal model gave
a very slightly better fit, with a reduced chi-square of $\chi^2_r = 1.06$,
as compared to the power-law model which had $\chi^2_r = 1.09$.  The
$\chi^2$-test indicates that both models are acceptable.  The difference
becomes even less significant if one combines the fits to zones 2--5.


\section{Discussion} \label{sec:disc}

With the new XMM data on the regions around the northwest radio relic in
Abell~3667, we have found a discontinuity in the X-ray surface brightness at
the outer edge of the radio relic. This increase in the X-ray surface
brightness inside of the relic compared to outside also corresponds to an
abrupt hardening in the X-ray spectrum. If the X-ray emission from the relic
is thermal emission from intracluster gas, then this feature corresponds to
a jump in the density, temperature, pressure, and specific entropy of the
gas.  The most likely origin of such a jump would be a merger shock in the
cluster. 

One can estimate the shock Mach number ${\cal{M}}$ from the compression behind the shock.
 From the X-ray surface brightness on either side of the surface brightness discontinuity, we estimate that the pre-shock and
post-shock electron densities are
$n_{e1} = (6.81 \pm 0.55) \times 10^{-5}$ cm$^{-3}$ and
$n_{e,2} = (1.32 \pm 0.08) \times 10^{-4}$ cm$^{-3}$, respectively.
This gives a shock compression of
$C \equiv n_{e,2} / n_{e1} = 1.94 \pm 0.19$, where the subscripts 1 and 2 refer to pre-shock and post-shock gas.
For a $\gamma = 5/3$ shock, the shock jump conditions give
\begin{equation}
\frac{1}{C} = \frac{3}{4 {\cal{M}}^2} + \frac{1}{4} \, .
\label{eq:shock_c}
\end{equation}
This gives a shock Mach number of ${\cal{M}} = 1.68 \pm 0.16$.  The effects
of projection, the finite width of the regions used to determine the surface
brightness, the XMM PSF, and the likely curvature of the shock front mean
that this is probably an underestimate.

One can also estimate the Mach number from the temperature jump at the shock
front.  Using the values for zones 10 and 4 for the pre-shock and post-shock
gas, we have a temperature jump of $ T_2 / T_1 = 2.68 \pm 1.19$.  For a
$\gamma = 5/3$ shock, the jump conditions give
\begin{equation}
\frac{T_2}{T_1} = \frac{5 {\cal{M}}^4 + 14 {\cal{M}}^2 - 3}{16 {\cal{M}}^2} \, .
\label{eq:shock_t}
\end{equation}
Using the observed temperature jump and propagating the errors gives
${\cal{M}} = 2.43 \pm 0.77$.
Because of the larger errors in the temperature, the uncertainty in this value is large.

If one takes the weighted mean of the Mach number from the shock compression
and shock temperature front, one finds $ {\cal{M}} = 1.71 \pm 0.16$.  Using
this value, the shock compression is $C = 1.97 \pm 0.19$.  The pre-shock
sound speed is $c_{s1} = 710 \pm 110$ km s$^{-1}$ which gives a shock speed
of $v_s = {\cal{M}} c_{s1} = 1210 \pm 220$ km s$^{-1}$.  Merger shocks of
similar and slightly higher strength has been detected in the bullet cluster
and A520 (Markevitch et al. 2002; 2005). However, the location of the shocks
in these clusters is in the core region, while in A3667 the location of the
shock is in the low-density region. Perhaps, the only other cluster where
the shock has been found in the low-density region is A754 (Krivonos et al.
2003; Henry et al. 2004).

Given the very low gas densities in this region roughly 2.2 Mpc from the
center of the cluster, and the high shock speed, it is possible that the
post-shock gas has not had time to come into electron-ion equipartition by
Coulomb collisions, or into collisional ionization equilibrium.
If the electrons are not strongly heated by the shock, the time scale for
the electron and ion temperatures to come into equipartition is
approximately
\citep{FL97,WS09}
\begin{equation}
t_{\rm eq} \approx 6.3 \times 10^8
\left( \frac{T_e}{10^7 \, {\rm K}} \right)^{3/2}
\left( \frac{n_p}{10^{-5} \, {\rm cm}^{-3} } \right)^{-1}
\left( \frac{\ln \Lambda}{40} \right)^{-1}
\, {\rm yr}
\, ,
\label{eq:teq}
\end{equation}
where $T_e$ is the electron temperature, $n_p$ is the proton density, and
$\ln \Lambda$ is the Coulomb logarithm.  Because the possible merger shock
is weak (has a low Mach number), the electrons will be heated significantly
by adiabatic compression, even if there is no shock electron heating.  For
the adapted value for the shock compression, adiabatic heating gives
$(T_{e2} / T_{e1} ) \approx 1.57$, while the full shock heating including
adiabatic compression gives $(T_{2} / T_{1} ) \approx 1.72$.  If the adapted
post-shock density and the post-shock electron temperature assuming only
adiabatic heating are used in equation~(\ref{eq:teq}), the approximate time
to reach equipartition is $t_{\rm eq} \approx 3 \times 10^8$ yr.  The speed
of the post-shock material relative to the shock is $v_2 = v_s / C \approx
610$ km s$^{-1}$.  Thus, the thickness of the region with a lowered electron
temperature would be $d_{\rm eq} \approx v_2 t_{\rm eq} \approx 0.2$ Mpc,
corresponding to an angular scale of $\theta_{\rm eq} \approx 3 \cos \phi$
arcmin.  Here, where $\phi$ is the angle between the central shock normal
and the plane of the sky.  This is comparable to the width of our spectral
zone 4.  So, this would argue that nonequipartition could be quite
important.  On the other hand, the apparent post-shock temperature is
actually a bit higher than expected assuming full shock heating and the
shock compression derived from the jump in X-ray surface brightness.  This
may indicate that the electrons are heated effectively at the shock.

At the observed post-shock temperature, one would expect the gas to achieve
ionization equilibrium after a time $t_{\rm ioneq} \sim 3 \times 10^{12} /
n_e$ sec \citep[e.g.,][]{Fuj+08}.  For the post-shock electron density, this
would give $t_{\rm ioneq} \sim 7 \times 10^8$ yr, which is even longer than
the timescale to reach equipartition.  However, the effects of
nonequilibrium ionization may be difficult to discern without much better
spectra than we currently have.

One argument favoring a thermal shock model for the jump in brightness and
hardness of the X-rays at the outer edge of the relic is that a similar jump
is seen in nearby regions where the radio emission from the relic is not
bright.  For example, the zones which are closest to the apparent shock
position but in regions with weak radio emission are zone 2 for the shocked
and zone 13 for the preshock. The density jump between these two zones
corresponds to a Mach number of $ {\cal{M}} \approx 2.2$.  The temperature
jump is basically identical to that measured on the radio relic, so the
estimate of Mach number from the temperature jump would not change.  Thus,
the jump in X-ray properties is similar in regions with bright or faint
radio emission, which is consistent with thermal emission from a shock, but
perhaps not with IC emission from the relic.

Could this merger shock be accelerating or re-accelerating the relativistic
electrons in the NW radio relic?  We start by estimating the energy
dissipated in the shock.  As a simple estimate of this, we take the change
in the kinetic energy flux across the shock.  This is given by
\begin{equation}
\Delta F_{\rm KE} = \frac{1}{2} \rho_1 v_s^3 \left( 1 - \frac{1}{C^2} \right) \, ,
\label{eq:shock_e}
\end{equation}
where $\rho_1$ is the pre-shock mass density in the gas.  Using our derived
values for the shock properties gives $\Delta F_{\rm KE} \approx 9 \times
10^{-5} \, {\rm erg} \, {\rm cm}^{-2} \, {\rm s}^{-1}$.  The width of the
relic from northeast to southwest is roughly 26\farcm3 or 1.63 Mpc.  Taking
the area of the shock perpendicular to the flow as a circle with this
diameter gives a perpendicular area of 2.09 Mpc$^2$.  With this size, the
total rate of conversion of shock kinetic energy is
\begin{equation}
\frac{d E_{\rm KE}}{dt} \approx 1.8 \times 10^{45} \, {\rm erg} \, {\rm s}^{-1} \, .
\label{eq:shock_dedt}
\end{equation}

We now compare the rate dissipation in the shock to the energy required to
power the radio source.  Given the steep radio spectrum of the relic, it
appears that the relativistic electrons are losing energy due to radiation.
We first consider the losses due to radio synchrotron emission.  Assuming
the relic is in steady-state, with the shock accelerating new relativistic
electrons and these particles losing energy by radiation, the total rate of
energy deposition in the electrons is given by the radio luminosity.
Unfortunately, there are few recent radio observations of the relic, and
some of the existing older values disagree, suggesting that some flux in
some observations may be missing due to the large extent and low surface
brightness of the relic, or that some of the fluxes include other sources.
The flux at 1.4 GHz is 3.7 Jy \citep{Joh-Hol04}, and the overall spectral
index from 85.5 MHz to 2.4 GHz is $\alpha \approx -1.1$ \citep{RWH+97}.  The
spectral index appears to steepen to $\alpha \approx -1.9$ from 1.4 GHz to
2.4 GHz, and to flatten below 1.4 GHz.  Thus, we approximate the radio
spectrum as
\begin{equation}
S_\nu  \approx 3.7 \, {\rm Jy} \, \left\{
\begin{array}{ll}
\left( \frac{\nu}{1.4 \, {\rm GHz}} \right)^{-0.9} & \nu \le   1.4 \, {\rm GHz} \\
\left( \frac{\nu}{1.4 \, {\rm GHz}} \right)^{-1.9} & \nu \ge 1.4 \, {\rm GHz}
\end{array}
\right.
 \, .
\label{eq:radio_spectrum}
\end{equation}
Then, the total radio luminosity of the relic is $L_{\rm radio} \approx 3.8
\times 10^{42}$ erg s$^{-1}$.  The IC luminosity of the relic is given by
$L_{\rm IC} = L_{\rm radio} ( 3.6 \, \mu{\rm G} / B )^2$, so the total
nonthermal luminosity is $L_{\rm NT} \approx 3.8 \times 10^{42} [ ( 3.6 \,
\mu{\rm G} / B )^2 + 1 ]$ erg s$^{-1}$.  If this energy is provided by the
shock acceleration of electrons, then the efficiency of shock acceleration
is
\begin{equation}
\epsilon \equiv \frac{\frac{d E_{e}}{dt}}{\frac{d E_{\rm KE}}{dt}}
\approx \frac{L_{\rm NT}}{\frac{d E_{\rm KE}}{dt}}
\approx 0.0021 \left[ \left( \frac{3.6 \, \mu{\rm G}}{B} \right)^2 + 1 \right]
\,  .
\label{eq:shock_eff}
\end{equation}
Below, we show that the magnetic field is the relic is at least $ B \ga 3$
$\mu$G, so that the correction factor in equation~(\ref{eq:shock_eff}) is
between one and 2.5.
Note that this acceleration efficiency is about one order of magnitude
smaller than the values of a few percent usually inferred from the radio
emission by Galactic supernova remnants \citep[e.g.,][]{RB07}.  This might
be due to the relatively low Mach number of this shock compared to these in
supernova remnants.  Alternatively, it may be that part of the X-ray
discontinuity at outer edge of the radio relic is due to IC emission, and
the shock is actually weaker than our estimate.

First order Fermi acceleration gives relativistic electrons with a power-law
spectrum $n(E) \, d E \propto E^{-p} \, d E$, where the power-law index is
$p = ( C + 2 ) / ( C - 1)$ and $C$ is the shock compression.  The spectral
index for radio emission near the shock should be $\alpha = - ( p - 1 ) /
2$, while the integrated radio spectrum of the relic and spectrum of hard
X-ray inverse Compton emission should be one power steeper. Using the value
of the compression determined above, this would give $\alpha \approx -1.55$
at the shock, and $\alpha \approx -2.55$ for the integrated spectrum of the
radio relic.  The observed radio spectra are flatter, with a value of
$\alpha \approx -0.7$ near the outer edge of the relic, and $-1.9 \la \alpha
\la -1.1$ for the integrated spectrum, depending on the observed frequency
\citep{RWH+97,Joh-Hol04}. This might indicate that we have underestimated
the shock compression, or that the simplest shock acceleration model for the
relic does not apply.

The fact that the radio spectrum steepens with projected distance from the
outer edge of the relic \citep{RWH+97} , and that the relic fades away to
the southwest rather than ending abruptly \citep{RWH+97,Joh-Hol04} might be
explained if the relativistic electrons are accelerated at the shock, and
then undergo radiative losses as they are advected away from the shock.
If the radio spectrum steepens at a frequency
$\nu_b$, the radiative age for electrons which produce the break is
\citep{vdLP69}
\begin{eqnarray}
t_{\rm rad} & \approx & 1.3 \times 10^8
\left( \frac{\nu_b}{1.4 \, {\rm GHz}} \right)^{-1/2}
\left( \frac{B}{3 \, \mu{\rm G}} \right)^{-3/2} \nonumber \\
& & \qquad \left[ \left( \frac{3.6 \, \mu{\rm G}}{B} \right)^2 + 1 \right]^{-1}
\, {\rm yr}
\, .
\label{eq:trad}
\end{eqnarray}
The speed of the post-shock material relative to the shock is $v_2 = v_s / C
\approx 610$ km s$^{-1}$.  Thus, the relativistic electrons will have moved
a distance $d_{\rm rad}$ away from the shock, where
\begin{eqnarray}
d_{\rm rad} & \approx & 0.082
\left( \frac{\nu_b}{1.4 \, {\rm GHz}} \right)^{-1/2}
\left( \frac{B}{3 \, \mu{\rm G}} \right)^{-3/2} \nonumber \\
& & \qquad
\left[ \left( \frac{3.6 \, \mu{\rm G}}{B} \right)^2 + 1 \right]^{-1}
\, {\rm Mpc}
\, .
\label{eq:drad}
\end{eqnarray}
The corresponding angular distance is
\begin{eqnarray}
\theta_{\rm rad} & \approx & 1\farcm3
\left( \frac{\nu_b}{1.4 \, {\rm GHz}} \right)^{-1/2}
\left( \frac{B}{3 \, \mu{\rm G}} \right)^{-3/2} \nonumber \\
& & \qquad
\left[ \left( \frac{3.6 \, \mu{\rm G}}{B} \right)^2 + 1 \right]^{-1}
\cos \phi
\,
\label{eq:thetarad}
\end{eqnarray}
where $\phi$ is the angle between the central shock normal and the plane of
the sky. This is roughly consistent with the observed thickness of the layer
at the front outer edge of the radio relic where the spectrum between 20 and
13 cm is observed to steepen dramatically \citep{RWH+97}.  On the other
hand, the full width of the relic is about 8\arcmin\ from NW to SE at 1.4
GHz.  Since this corresponds to several electron radiation loss lengths,
this may indicate that there is re-acceleration of particles within the
relic, perhaps by turbulence produced by the passage of the merger shock.
Alternatively, far from the NW edge of the relic, we may be seeing radio
emission from relativistic electrons which have been recently accelerated
and which are located at the front or back edge of a convex shock region .

This argument neglected the diffusion of the relativistic electrons.
If Bohm diffusion applies, the diffusion coefficient is
\citep{BBP97},
\begin{equation}
D(E) \approx 2 \times 10^{29}
\left( \frac{E}{1 \, {\rm GeV}} \right)^{1/3}
\left( \frac{B}{1 \, \mu{\rm G}} \right)^{-1/3}
\, {\rm cm}^2 \, {\rm s}^{-1}
\, .
\label{eq:bohm}
\end{equation}
The electrons which produce the radio emission at 20 cm have $E \approx 10 (
B / 3 \, \mu{\rm G} )^{-1/2}$ GeV, which leads to $D(E) \approx 3 \times
10^{29} ( B / 3 \, \mu{\rm G} )^{-1/2}$ cm$^{2}$ s$^{-1}$.  During the
radiative loss time $t_{\rm rad}$, these electrons would diffuse a distance
of $d_{\rm diff} \approx [ D(E) t_{\rm rad} ]^{1/2} \sim 0.01$ Mpc, which is
smaller than the advection distance.  We also note that the sharp outer edge
of the radio relic suggests that diffusion is not very important on the
scale of the size of the relic.

Alternatively, the increase in the X-ray brightness and hardening of the
spectrum at the outer edge of the relic might be the result of inverse
Compton emission, rather than thermal emission from the shocked gas.  As
note in \S~ \ref{sec:spectra}, the signal-to-noise of the X-ray spectra do
not allow these two models to be distinguished.  If the jump in X-ray
emission and hardness is due to inverse Compton emission, one would still
need to explain why the relic is located where it is.  That is, how are the
relativistic electrons in the relic generated?  It might be that the
particles are due to first order Fermi re(acceleration) from a merger shock;
in this case, there would still need to be a shock at the front of the
relic, but the X-ray emission in this region might be predominantly
nonthermal.  It is also possible that the electrons are accelerated by some
other mechanism.

Taking the nonthermal fits as an upper limit on the true flux of inverse
Compton emission coming from the relic in regions 4 and 5, we can derive a
lower limit on the average magnetic field strength $B$ as shown by
\citet{HR74}. Considering monochromatic fluxes $F_R(\nu_R)$, $F_X(\nu_X)$
at frequencies $\nu_R$, $\nu_X$, for a power-law distribution of electrons,
the expression for the magnetic field is
\begin{eqnarray}
B & =  & C(p) (1+z)^{(p+5)/(p+1)} \left(\frac{F_R}{F_X}\right)^{2/(p+1)} \nonumber \\
& & \qquad
   \left(\frac{\nu_R}{\nu_X}\right)^{(p-1)/(p+1)}
\, .
\label{eq:bexact}
\end{eqnarray}
The value of the proportionality constant
$C(p)$ can be found from the ratio of the synchrotron flux
\citep[][eqn.~18.49]{Lon94} to the IC flux \citep[][eqn.~7.31]{RL79}. For
region 4 we use the photon index found for region 5, since this is more
consistent with the spectral index of the radio relic. The lower photon
index found in region 4 may be the result of coincident thermal emission,
and therefore our estimate of $B$ would be low.  Assuming the X-ray flux in
these regions is nonthermal with a photon index in the range $1.9 \lesssim
\Gamma \lesssim 2$, we find magnetic field strengths in the range $2.3 \,
\mu{\rm G} \lesssim B \lesssim 4.5\, \mu{\rm G}$.  Since it is unknown what
fraction of the emission is thermal, $B$ may in fact be larger.  Therefore
we estimate a lower limit for the magnetic field at the location of the
radio relic of $B > 3 \, \mu$G.  This limit is just consistent with Faraday
rotation measure estimates from the the observation of background radio
galaxies in the zone close to the relic \citep{Joh-Hol04}.

\section{Conclusions}

We have made a new X-ray observation of the northwest radio relic region in
the Abell 3667 cluster in order to gain insight on both thermal and
non-thermal phenomena that should be associated with it.  We detect a jump
in both the surface brightness and the hardness of the X-ray emission at the
outer edge of the relic.  The X-ray spectra of these regions is consistent
with the extra emission on the relic being either thermal emission from a
merger shock or inverse Compton emission from relativistic electrons in the
radio relic.  There are some arguments that favor the shock model, which
implies a shock with a Mach number of ${\cal M} \approx 2$ and a shock speed
of $\sim$1200 km s$^{-1}$.  The energy content of the relativistic particles
in the radio relic can be explained if they are (re)-accelerated by the
shock with an efficiency of $\sim$0.2\%.  The merger shock model and shock
acceleration of the relativistic electrons can explain, at least roughly,
the radio surface brightness and spectral distributions in the relic.

Alternatively, the jump in the brightness and hardness of the X-rays at the
outer edge of the relic might be due to inverse Compton emission from the
relic. Assigning all the emission from the relic to the IC component, we
derive a lower limit on the magnetic field in the relic of $\ga$3 $\mu$G.

These observations have yielded the first robust characterization of the
X-ray properties near the northwest radio relic and possible merger shock in
Abell 3667. Unfortunately, the data are not sufficiently deep to determine
if the relative roles the thermal shock emission and nonthermal IC emission.
Recently, we were approved for a very long observation of this regions as an
XMM-Newton Large Program.  These data should allow us to determine the
efficiency of particle acceleration and the relative roles of thermal and
nonthermal processes in this merger shock.

\acknowledgments The work was supported by NASA primarily through XMM-Newton
Grant NNX08AZ34G, but also through Suzaku Grants NNX06AI37G, NNX08AZ99G, and
NNX09AH74G.  AF thanks UVA for the hospitality during his visits. AF thanks
Francesco Miniati and Melanie Johnston-Hollitt for useful discussions. We
thank the referee for the constructive comments, which improved the quality
of the presentation of this paper. Basic research in radio astronomy at the
Naval Research Laboratory is supported by 6.1 Base funding.

{\it Facilities:} \facility{XMM-Newton}, \facility{Suzaku}




\end{document}